\documentclass[conference]{IEEEtran}
\IEEEoverridecommandlockouts
\usepackage{cite}
\usepackage{amsmath,amssymb,amsfonts}
\usepackage{algorithmic}
\usepackage{graphicx}
\usepackage{textcomp}
\usepackage{gensymb}
\def\BibTeX{{\rm B\kern-.05em{\sc i\kern-.025em b}\kern-.08em
    T\kern-.1667em\lower.7ex\hbox{E}\kern-.125emX}}
    
\begin{document}
\bstctlcite{IEEEexample:BSTcontrol}

\title{Privacy-Preserving Social Distancing Bracelet\\
\thanks{The authors would like to express their sincere gratitude to the Deanship of Scientific Research at KFUPM for the immense support provided. Our gratitude is extended to the Technology Advancement and Prototyping Centre (TAPC) at the Dhahran Techno Valley for the technical support and facilities.}
}

\author{\IEEEauthorblockN{Asaad AlGhamdi}
\IEEEauthorblockA{\textit{Information \& Computer Science Department} \\
\textit{King Fahd University of Petroleum and Minerals}\\
Dhahran, Saudi Arabia \\
s201516850@kfupm.edu.sa}
\and
\IEEEauthorblockN{Louai Al-Awami}
\IEEEauthorblockA{\textit{Computer Engineering Department} \\
\textit{King Fahd University of Petroleum and Minerals}\\
Dhahran, Saudi Arabia \\
louai@kfupm.edu.sa}

}

\maketitle

\begin{abstract}
This demo presents a functional Proof-of-Concept prototype of a smart bracelet that utilizes IoT and ML to help in the effort to contain pandemics such as COVID-19. The designed smart bracelet aids people to navigate life safely by monitoring health signs; and detecting and alerting people when they violate social distancing regulations. In addition, the bracelet communicates with similar bracelets to keep track of recent contacts. Using RFID technology, the  bracelet helps in automating access control to premises such as workplaces. All this is achieved while preserving the privacy of the users.
\end{abstract}


\section{Introduction}
Technology can play a crucial rule at times of crises. This is particularly true in the current intricate case of the COVID-19 pandemic where the challenge is of perplexing nature. Technologies such as the Internet-of-Things (IoT), Artificial Intelligence (AI), Big Data, and Robotics, to mention a few, can provide valuable insights and reduce dependence on human actors leading to a safer and faster containment.

Contact tracing aims to identify potentially infected individuals in a timely fashion to properly treat and isolate the spread of infections. While contact tracing has long been in practice, it has - until recently - been manually performed. Many automated contact tracing protocols have been recently proposed in the literature such Pan-European Privacy-Preserving Proximity Tracing (PEPP-PT) \cite{pepp202}, Coalition \cite{Loiseau2020}, and the Decentralized Privacy-Preserving Proximity Tracing (DP-3T) \cite{Troncoso2020}. Compared to other protocols, (DP-3T) has been more successful where privacy is a key concern. In DP-3T, devices regularly broadcast unique and continuously-changing tags. Each device stores both the set of tags it broadcasted and the tags it heard. When a user is tested positive, he/she can voluntarily permit the device to send the tags it broadcasted to a central database that stores all “infected” tags. Other devices regularly check and compare the tags they heard against the database. If a match is found, the device alerts the user to take proper measures. The main strength of DP-3T is that no information that can be used to identify the user is collected, hence, providing strict privacy. However, the protocol lacks a mechanism to track duration of exposure when tags change. In this work, we use the DP-3T protocol due to its strict privacy measures. Our implementation in contrast adds an extension to enable tracing the duration of exposure even when the tags change without breaking the privacy promise. Our system also uploads contacts automatically when infection symptoms are detected which improves the reaction time of the system. 

In this context, we have successfully developed an IoT-based solution to enable automatic tracing of potential COVID-19 cases. The solution is comprised of a smart bracelets equipped with sensors to monitor the vital signs of its wearer and provide an early warning when abnormal symptoms arise. In addition, the bracelet can detect when another bracelet is physically too close and reminds the user to respect social distancing regulations. The bracelet also listens and stores messages form nearby bracelets which are then used to warn those who had contacted a confirmed COVID-19 patient; in order to take necessary measures such as executing a quarantine. As privacy is a key concern in such applications, we have adopted a fully privacy-preserving approach. Yet, if necessary, the bracelet can also be used by authorities and employers to control access to certain places. Using Radio Frequency IDentification (RFID), the bracelet can be scanned using an electronic reader to determine the users' risk level, and consequently grant or deny him/her access.

\section{Research Components and Issues}

The presented application combines a number of interesting research issues. The first is the privacy preservation while maintaing the ability to track the identity of the subjects and the duration of exposure. The second issue is the proximity detection and computation of separation distance using RSSI. As RSSI is typically unreliable, machine learning was used to build a model to perform this task. The third is chrachterising couhging and health condition through fusion of sensor data. In addition to presenting exciting research issues, the application demonistrates a well-rounded example of integration between many interrelated technologies, namely IoT, ML, Cloud Computing, RFID, and BLE; working hand-in-hand to deliver a timely and crucial application. The demo includes a functional prototype demonistrating all the described functionalites working toegther to achieve the claimed application.

\section{Design \& Implementation}
\subsection{System Model}
\begin{figure}[t!]
\centerline{\includegraphics[scale=0.43]{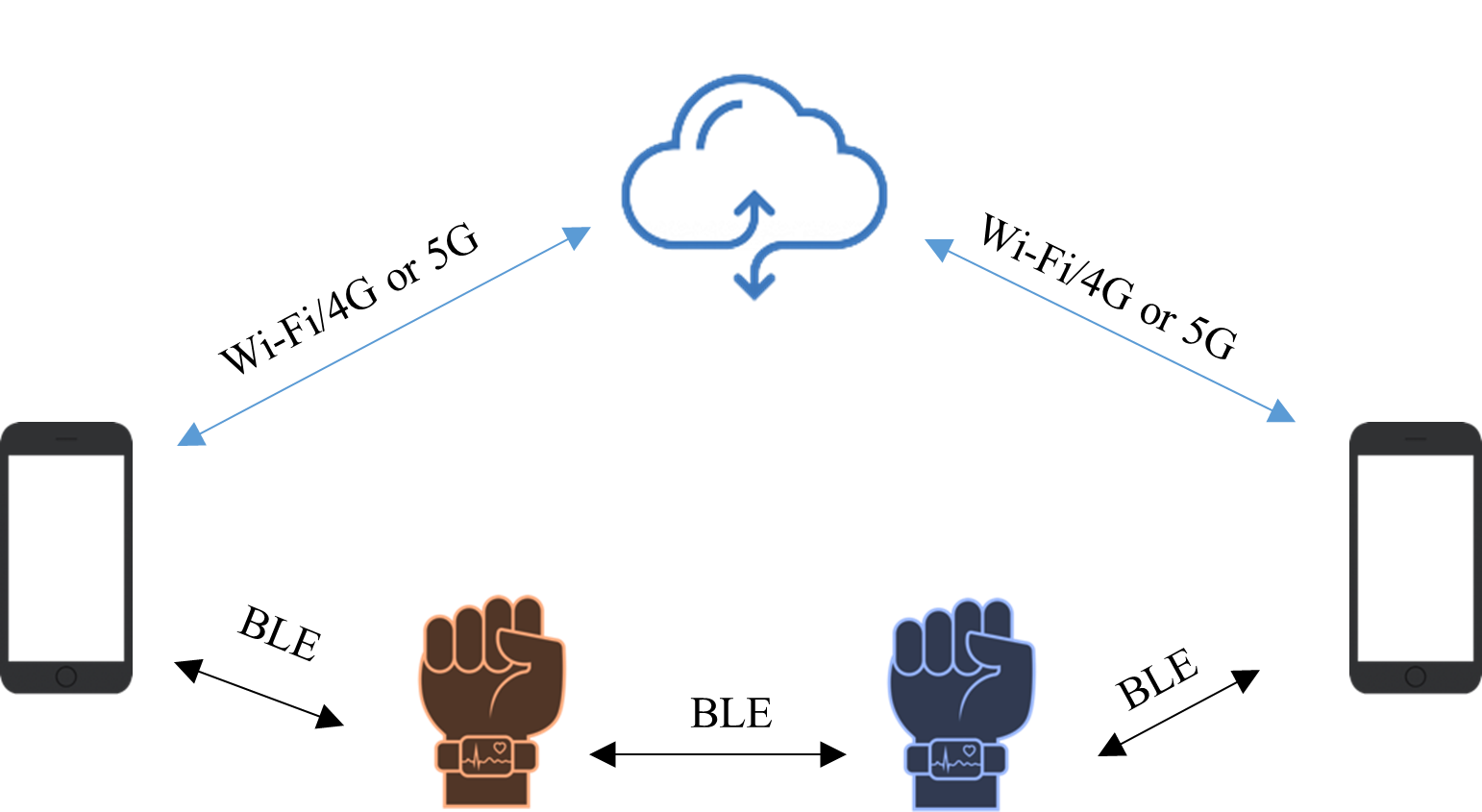}}
\caption{Smart Bracelet System Architecture.}
\label{fig:model}
\end{figure}
As shown in Figure \ref{fig:model}, the system is comprised of a number of smart bracelets each is equipped with a Bluetooth Low-Energy (BLE) and a sensing modules. The BLE module provides a beaconing service and enables pairing the bracelet to a smart phone. On the other hand, the sensing module encompasses a number of sensors used to monitor the health status of the wearer. The system also relies on a cloud service that is reachable over the Internet to test tags.

\subsection{Health Monitoring}
\begin{figure}[t!]
\centerline{\includegraphics[scale=0.37]{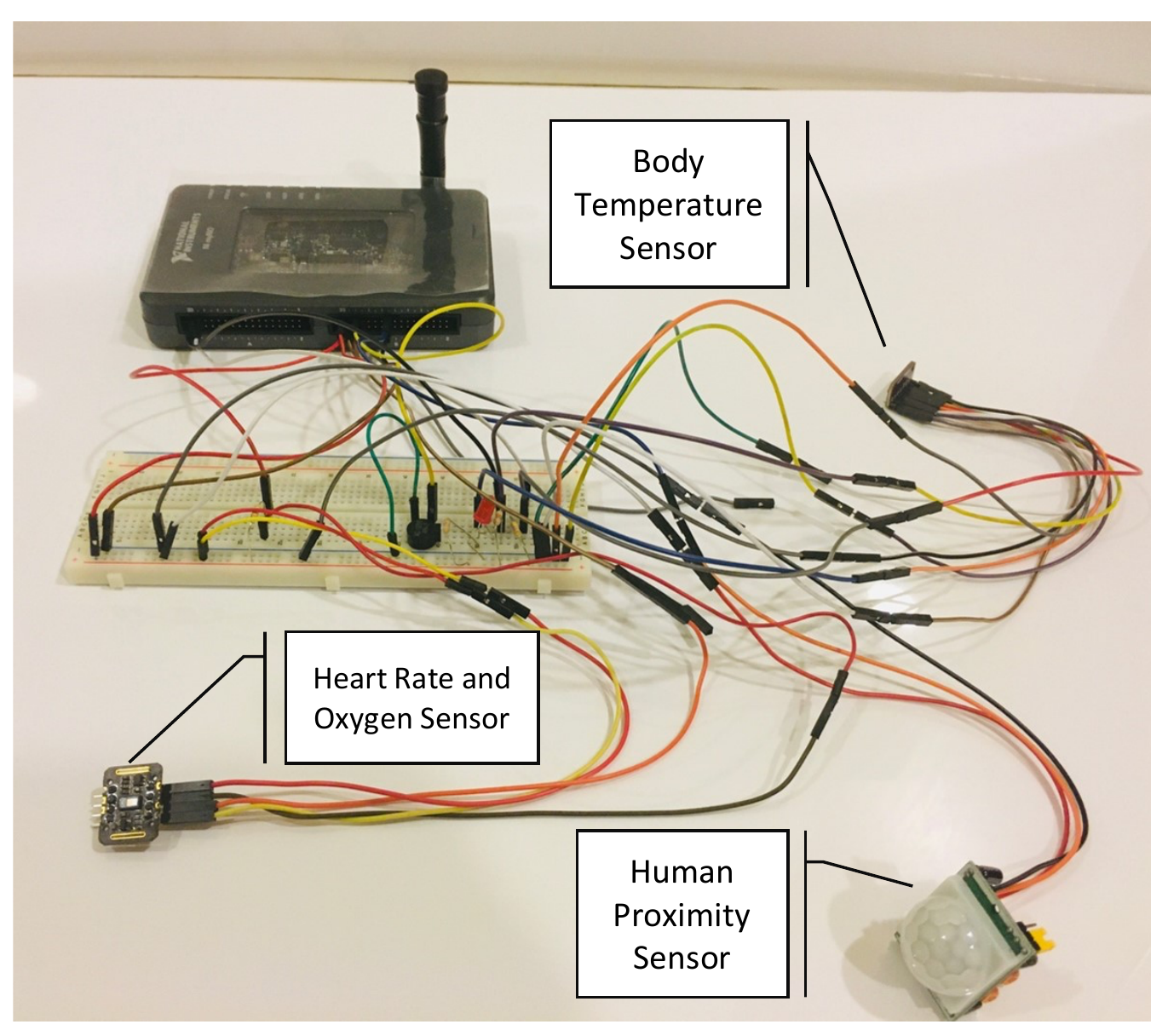}}
\caption{Health Monitoring Module.}
\label{fig:health-module}
\end{figure}

The role of the health monitor is to track readings from various sensors to infer the health status of the user. The module also detects when the bracelet wearer coughs, or when another person is too close to the bracelet. The module can be seen in Figure \ref{fig:health-module}. User risk level can be either: 1) No-risk: a person who has no symptoms, no excessive violations, and no infected contacts 2) Low-risk: a person who had either been in contact with an infected person or have excessive social distancing violations 3) High-risk: a person with abnormal health symptoms, or who has been in contact with an infected case and has excessive social distancing violations.

We define abnormal health symptoms as follows: 1) Detection of high temperature (Fever greater than 38.0 \degree{C}) 2) Excessive coughing 3) Blood Oxygen less than 90\%. Once any of the aforementioned symptoms is detected, the user is classified as a high-risk individual. Temperature and Blood Oxygen readings are sampled periodically using medical grade external sensors. Social distancing violations and coughing are detected through the combination of general-purpose sensors. 

\subsection{Contact Tracing}
\begin{figure}[t!]
\centerline{\includegraphics[scale=0.45]{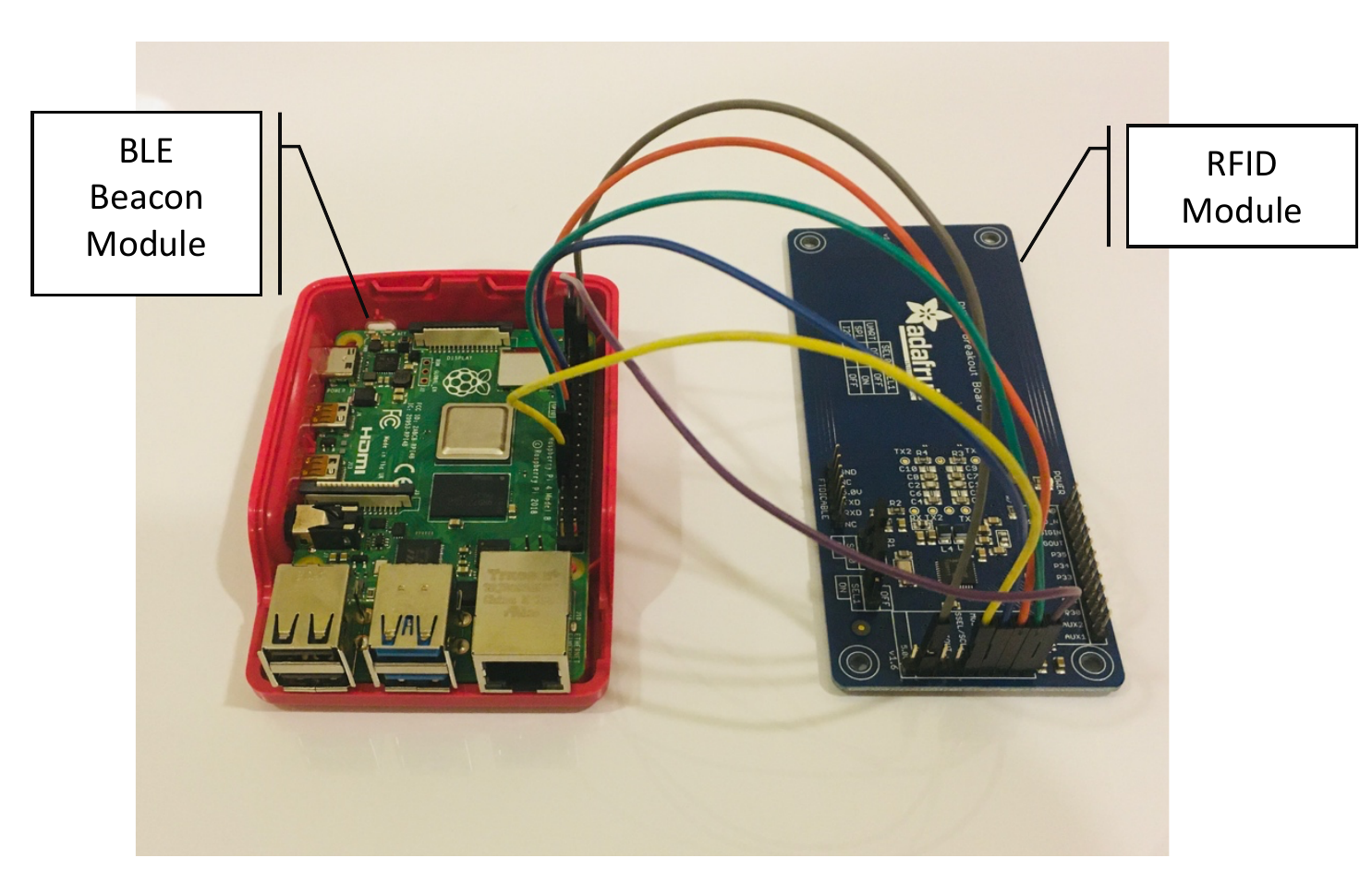}}
\caption{Contact Tracing Module.}
\label{fig:contact-tracing}
\end{figure}
The contact tracing module frequently sends BLE beacons while listening to beacons coming from other bracelets. The module alerts the user when violating social distancing by inferring the distance form the RSSI of the received beacon. We developed a supervised machine learning model to estimate the distance based on the transmitted power and the RSSI. When the risk level of the wearer changes, the module - if permitted by the user - uploads the user's generated tags to the cloud and stores the new risk level on the RFID tag. This module has been implemented using Raspberry Pi as seen in Figure \ref{fig:contact-tracing}. The software of the module includes multiple threads for various tasks. The software maintains two tables: \textbf{contacts} and \textbf{tags}. The \textbf{contacts} table contains all the heard tags, the time each tag was first heard, and the total exposure time; while the \textbf{tags} table stores the tags sent by the node itself.

\subsection{Privacy Preservation}
To preserve privacy, each device uses a random BLE address when sending beacons. In addition, devices generate random tags generated using a secure hash function to avoid collisions. Both BLE addresses and tags change every $15$ minutes. When devices hear beacons from a nearby device, the random tags heard are stored locally in addition to the duration of exposure. When a user is deemed infected, the tags he used are uploaded to the cloud. Other devices constantly contact the cloud to retreive the list of infected tags and compare them to the locally stored ones. This way, exposure detection is identified by the end devices rather than the cloud which improves privacy.
\subsection{Cloud Service}
The cloud service includes a database holding all the infected tags and two main server-side scripts: one to receive and store tags in the database and the other to process and compare overheard tags against infected tags. When a bracelet wants to test its contacts, it uploads the tags to the cloud. The service then checks if any of the contacts is in the infected tags database. Infected tags coming from the same node are linked together. This allows computing the exposure duration across different tags coming from the same bracelet. If an exposure with an infected contact for 15 minutes or more is found, the scripts returns a positive response implying that the user may be infected and needs to take proper measures. 

\subsection{Access Control}
The purpose of the access control module is to enable scanning the health status of the user using RFID technology. This module includes a physical RFID tag that can be programmed by software to reflect the value of the risk level of the user. The module can be used in conjunction with an RFID reader to automate access control at work, or public places if required.

\section{Conclusion}
This paper demonstrates an important application of e-Health using IoT. Such an application can play a crucial role in controlling the spread of vicious diseases such as COVID-19. While a standard application of contact tracing can easily be implemented using off-the-shelf technologies, this specific target application combines a number of challenging aspects: 1) Privacy preservation 2) Exposure duration determination 3) Physical distance estimation 4) Health status determination. The prototype demonstrates how an interesting mix of technologies such as IoT, RFID, ML, BLE, and localization can all be combined to achieve effective applications.


\bibliographystyle{IEEEtran}
\bibliography{references}

\end{document}